\documentclass[pra,aps,amsfonts,groupedaddress,showpacs,showkeys]{revtex4-1}
\usepackage{epsfig,tabularx,array,booktabs,calc,multirow,amsmath,color}
\usepackage{bm}

\def\shiftdown#1{#1\llap{\lower.04ex\hbox{#1}}}

\newcommand{\beaa}{\begin{eqnarray*}} 
\newcommand{\enaa}{\end{eqnarray*}}
\newcommand{\bea}{\begin{eqnarray}}
\newcommand{\ena}{\end{eqnarray}}
\newcommand{\be}{\begin{eqnarray}} 
\newcommand{\eq}{\begin{eqnarray}} 
\newcommand{\en}{\end{eqnarray}}

\begin{document}

\title{Hidden-charm and radiative decays of the $Z(4430)$ as a hadronic $D_1 \overline{D^\ast}$ bound state} 

\author{
Tanja Branz, 
Thomas Gutsche, 
Valery E. Lyubovitskij
\footnote{On leave of absence from the
Department of Physics, Tomsk State University,
634050 Tomsk, Russia}
\vspace*{1.2\baselineskip}}

\affiliation{Institut f\"ur Theoretische Physik,
Universit\"at T\"ubingen,
\\ Auf der Morgenstelle 14, D-72076 T\"ubingen, Germany
\vspace*{0.3\baselineskip}\\}

\date{\today}

\begin{abstract} 

We study the $Z(4430)^\pm$ meson within a heavy hadron 
molecule interpretation where the $Z$ is considered as a bound state of a vector $D^\ast(2010)$ and an axial-vector $D_1(2420)$. We give predictions for 
the strong hidden-charm $Z(4430)^\pm \to \pi^\pm \psi$ and 
$\pi^\pm\psi^\prime$ decay widths and also study 
the radiative $Z^\pm\;(J^P=1^-) \to \pi^\pm\gamma$ decay properties 
in a phenomenological Lagrangian approach. Our findings are qualitatively in line with the experimental observation that the $\pi^\pm\psi^\prime$ transition dominates over 
the $\pi^\pm\psi$ decay mode despite a smaller phase space. The width for the radiative mode $\pi^\pm\gamma$ is sizable, allowing possible detection in future experiments.
\end{abstract}



\maketitle

\newpage

\section{Introduction}

The spectrum, production and decay properties of observed hadrons still pose major challenges
in their theoretical understanding. In particular, already in the light meson sector we
have experimentally observed resonances, as for instance the low-lying scalar mesons, which cannot be simply
and consistently explained by a leading order quark-antiquark ($q\bar q$) structure.
But an unique interpretation
of such anomalous meson resonances is also not at hand, not even for a single case.
Detailed discussions concerning the possible non-$q\bar q$ nature of meson resonances find their
repetition in the heavy meson and especially the charmonium sector~\cite{Godfrey:2008nc,Olsen:2009ys,*Olsen:2010zz}.
Here investigations are essentially fueled by the enormous progress
on the experimental side with the discovery of many new charmonium-like states by the Belle and {\it BABAR} Collaborations at the $B$-factories (see e.g. recent conference proceedings~\cite{Dunwoodie:2009zz,*Palano:2009zz,*Kreps:2009ne,*ChengPing:2009vu}), but also in experiments by the CDF, D0 and CLEO collaborations \cite{Yi:2010rv,*Mitchell:2007ms}.  
Properties of these newly observed $X,\,Y$ and $Z$ mesons
cannot be easily explained within the standard charm-anticharm ($c\bar c$) assignment.
For instance, besides the overpopulation in the $c\bar c$ spectrum due to the numerous occurrence
of $X$, $Y$ and $Z$ mesons decay patterns of these mesons are in some cases in contradiction to the
standard charmonium predictions.
Here one example is the sizable  $\omega \phi$ hidden charm decay width of the $Y(3940)$~\cite{Swanson:2006st,Eichten:2007qx,Branz:2009yt}
which should be significantly suppressed in the charmonium picture.
Some of these new charmonium-like states are considered good candidates for possessing a hadronic
substructure which goes beyond the standard $c\bar c$ assignment ranging from quark-gluon hybrid
mesons~\cite{Close:1995eu,Luo:2005zg} and tetraquark states~\cite{Liu:2008qx} to dynamically generated
states~\cite{Molina:2009ct} or bound states of two mesons called hadronic
molecules~\cite{Liu:2009ei,Branz:2009yt}. 
A review on the experimental situation with a first overview of the present theoretical understanding
is e.g. given in~\cite{Godfrey:2008nc,Olsen:2009ys,Swanson:2006st,Eichten:2007qx}.

The observation of the charged $Z(4430)^{\pm}$ by the Belle Collaboration~\cite{:2007wga}
presents so far the culmination of the series of newly discovered $X$, $Y$ and $Z$ mesons.
The $Z^+$ was originally found as a relatively narrow enhancement in
exclusive $B\to K\pi^+ \psi^\prime$ decays
with mass $m_Z=4433\pm4$(stat)$\pm2$(syst) MeV and width
$\Gamma_Z=45^{+18}_{-13}$(stat)$^{+30}_{-13}$(syst) MeV.
The {\it BABAR} Collaboration also searched for the $Z^+$ resonance in $B^+\to K \pi^+\psi^\prime$ decays
but without positive evidence for a narrow $Z^+\to \pi^+\psi^\prime$ signal~\cite{:2008nk}. They report a branching-fraction upper-limit for example for the process ${\cal B}(B^0\to Z(4430)^-K^+,\,Z^-\to\psi^\prime\pi^-)<3.1\cdot10^{-5}$ at the 95\% c.l. It was speculated that the resonant structure, observed by Belle, might arise from interference effects in the
$K\pi$ rather than in the $\pi \psi^\prime$ channel. After this report the Belle Collaboration~\cite{:2009da} reanalyzed their data sample in a full
Dalitz-plot formalism now with a 6.4 $\sigma$ signal for $Z^+\to \pi^+\psi^\prime$ with
mass $m_Z=4443^{+15+19}_{-12-13}$ MeV and width $\Gamma=107^{+86+76}_{-43-56}$ MeV.
This reanalysis confirms and supersedes the previous $Z^+$ resonance parameters of~\cite{:2007wga}.
The larger errors contain e.g. the uncertainties in the spin assignment of the $Z^+$ ($J=0,1$)
and in the orbital angular momentum in the $B$ decay. The reanalysis by Belle results in a product branching fraction ${\cal B}(\bar B^0\to K^- Z(4430)^+){\cal B}(Z(4430)^+\to \pi^+\psi^\prime)=(3.2^{+1.8+5.3}_{-0.9-1.6})\cdot10^{-5}$~\cite{:2009da} which is consistent with the {\it BABAR} upper limit. The decay mode $\psi^\prime \pi^+$, assuming
standard conservation laws, leads to an identification of the $Z^+$ as an isotriplet state with
positive $G$-parity.
The $J^P$ quantum numbers remain to be determined.
As in other cases of the $X,\,Y$ and $Z$ mesons, see e.g. the $Y(3940)$, the $Z^+$ also shows
a sizable coupling to the hidden charm decay channel. As a consequence the partial decay
width of $Z^+\to\pi^+\psi^\prime$ is expected to be on the MeV scale~\cite{Godfrey:2008nc}.
In comparison, for the conventional $c\bar c$ configuration open charm decay modes are dominant
whereas hidden charm decay channels are Okubo, Zweig and Iizuka (OZI)-suppressed and typically
result in decay widths of a few keV only~\cite{Eichten:2007qx,Swanson:2006st}.
Since this resonant structure was observed in the invariant mass of
$\pi^{\pm} \psi^{\prime}$ the $Z^{\pm}$ as a charmonium state with isospin  $I=1$ 
is a truly exotic resonance which might offer the possibility to uniquely pin down a non-$c\bar c$
structure.
A further key issue concerns the observation mode of the $Z^+$. While the $Z^+$ was seen by Belle in
the $\psi^{\prime}$ channel no signal was reported in the $\pi^+\psi $ mode implying
a large ratio $R={\Gamma(Z^+\to\pi^+\psi^\prime)}/{\Gamma(Z^+\to\pi^+\psi)}$. Note that the suppressed
mode $\pi^+\psi $ is favored by phase space. Although some arguments~\cite{Maiani:2008zz,Maiani:2007wz,Meng:2007fu,Rosner:2007mu} were put forward
to understand this dynamical selection rule, a full quantitative explanation is not given yet. The Belle Collaboration reported~\cite{Mizuk:2008me} on further charged states in the charmonium sector. There is evidence for two charged resonances in the $\pi^+\chi_{c1}$ channels termed $Z^+_1(4050)$ and $Z_2^+(4250)$. However, the signal is much poorer than for the $Z(4430)^+$.

Several non-$c\bar c$ structure interpretations have already been discussed in relation
to the $Z(4430)^\pm$ (since the $Z^{\pm}$ carries charge a $c\bar c$ hybrid configuration is
obviously excluded). 
The $Z^+$ is considered a candidate for a radially excited $c\bar c u\bar d$
tetraquark~\cite{Maiani:2008zz,Liu:2008qx,Maiani:2007wz,Li:2007bh}, for a less compact hadronic
meson molecule~\cite{Ding:2007ar,Cheung:2007wf} or be just due to
threshold effects~\cite{Rosner:2007mu}. Further explanations for the occurrence of the $Z(4430)^+$
are a cusp in the $D_1 D^{\ast} $ channel~\cite{Bugg:2008wu}, a radial excitation
of a $c\bar s$ configuration~\cite{Matsuki:2008gz}, a baryon-antibaryon (baryonium) bound
state~\cite{Qiao:2007ce} or even a $\psi^{\prime} $ bound state in mesonic matter 
\cite{Voloshin:2007dx}.
The molecular interpretation is rather natural since the
$Z(4430)^+$ mass lies extremely close to the $D_1(2420)\overline{D^\ast}$ threshold at 4.43 GeV.  
Assuming an $S$-wave $D_1(2420) \bar D^\ast$ bound state possible quantum numbers are
$J^{P}=(0,1,2)^{-}$. This would also imply that the $Z^+ \to \pi^+ \psi^{\prime}$ decay proceeds
in a final state $P$-wave as opposed to an $S$-wave for the $J^P =1^+$
tetraquark proposition. Note that the $D_1(2420)$ is fairly narrow with a total width
of about 20 MeV, which would principally allow the formation of a hadronic bound state.
Alternatively, the bound axial-vector charm meson can also be identified with the $D^{\prime}_1 \equiv
D_1 (2430)$ which within errors is degenerate in mass with the $D_1(2420)$ but has
a rather large width of about 400 MeV.

Aspects of the hadronic-molecule interpretation have been studied within various models including
meson-exchange potential approaches~\cite{Close:2010wq,Liu:2007bf,Liu:2008xz},
QCD sum rules~\cite{Lee:2007gs,Bracco:2008jj} and effective Lagrangian techniques~\cite{Liu:2008yy,Meng:2007fu}.
First analyses~\cite{Liu:2007bf,Close:2010wq} based on the long-range one-pion exchange (OPE) mechanism
conclude that the isovector $D_1 \overline{ D^\ast}$ or $D_1^{\prime} \overline{ D^\ast}$ systems cannot
form a $J^{PC}=0^{--}$ or $1^{--}$ bound state. Based on the binding energy of the $D_1\overline{ D^\ast}$
system of around a few MeV the OPE study in~\cite{Close:2010wq} suggests a probable isovector $1^{-+}$
assignment for the $Z^+$.
Further inclusion of sigma meson exchange leads to $S$-wave binding for $D_1^{\prime} \overline{D^\ast}$
with $J^{P}=0^-, ~1^-,~2^-$~\cite{Liu:2008xz}, but the large width of the $D_1^{\prime}$ probably
disfavors the formation of a molecular state.
Inclusion of the sigma-exchange potential also leads to binding for the $D_1 \overline{D^\ast}$
configuration but only for $J^P =0^-$ and at the price of a large cutoff which in turn leads
to an enhanced attraction of one-pion-exchange.
An evaluation in the context of QCD sum rules~\cite{Lee:2008tz} favors the molecular $D_1\overline{ D^\ast}$ bound state interpretation
of the $Z^+$ with quantum numbers $0^-$. The study of the low-energy $ D_1\overline{D^{\ast}}$ interaction
in a quenched lattice calculation also indicates attraction in the $J^P =0^-$ channel,
but this effect is considered probably to weak to lead to the formation of a bound state
\cite{Meng:2009qt}.

Two-body decays of the $Z^+$ were also analyzed in effective Lagrangian
methods. The open charm decays $D^+ \overline{ D^{\ast 0}} $, $D^{\ast +} \overline{ D^0}$ and $D^{\ast +}
\overline{ D^{\ast 0}}$ were analyzed in~\cite{Liu:2008yy} and, in spite of ill determined
coupling constants, argued to be suppressed
in the molecular interpretation while dominant for the tetraquark configuration.
More importantly, the $Z^+$ was observed in the hidden charm mode $ \pi^+ \psi^{\prime}$ while
no signal was seen in the $\pi^+\psi  $ decay channel. In the unpublished work of~\cite{Meng:2007fu}
these channels were investigated in addition to the dominant $D^{\ast}D^{\ast} \pi$ decay, but the conclusion
on the possible suppression of $J/\psi $ depends very much on form factors and the regularization
in the loop diagrams.

In the present work we reconsider and pursue a quantitative explanation of the hidden charm
decay modes $Z^+ \to \pi^+\psi $ and $\pi^+\psi^{\prime} $ in the context of
a molecular $D_1 \overline{D^\ast}$  bound state interpretation. In addition we determine the radiative
decay width $Z^+ \to \pi^+ \gamma$ as a further key feature of the molecular idea.
As suggested by the above mentioned studies related to possible binding 
we consider the quantum numbers $J^P =0^-$ and $1^-$ for the hadronic molecule.
In technical aspects we proceed in analogy to the open and
hidden-charm hadrons $D_{s0}^\ast(2317)$, $D_{s1}(2460)$, $X(3872)$, $Y(3940)$, $Y(4140)$, $\Lambda(2940)$, 
etc.~\cite{Branz:2009yt,Faessler:2007gv,Dong:2009yp,Dong:2009uf,Faessler:2007cu,*Faessler:2007us,%
*Dong:2008gb,*Dong:2009tg,*Dong:2010gu} considered
previously as hadronic molecules.
For our analysis we use an effective Lagrangian approach for the treatment of
composite objects --- molecular 
states~\cite{Branz:2009yt,Faessler:2007gv,Faessler:2007cu,Faessler:2007us,%
Dong:2008gb,Dong:2009yp,Dong:2009tg,Dong:2009uf,Dong:2010gu}. 
The hadronic bound state is set up by means of the compositeness
condition~\cite{Weinberg:1962hj,*Salam:1962ap,*Efimov:1993zg,*Ivanov:1996pz,*Ivanov:1996fj} 
which also allows for a self-consistent determination of
the coupling strength between the hadronic molecule and its constituents.

This work is organized as follows: In the following section we discuss the set up of the mesonic bound state and introduce the effective Lagrangian approach which we use to study the decay properties of hadronic bound states. In the subsequent sections we apply our method in order to compute the radiative $Z^+\to \pi^+\gamma$ decay in section \ref{sec:radiative} and strong $Z^+\to \pi^+\psi^{(\prime)}$  decays in section \ref{strong}. Our results are presented in section \ref{sec:results}. At the end of this work (in section \ref{sec:conclusions}) we give a short summary of our findings and draw the conclusions.

\section{Theoretical Approach}

In the present study we assume the $Z^+$ to be a pure bound state of an axial $D_1^{(\prime)}$ and a vector $D^\ast$ meson. In the charmed meson spectrum two nearby $P$-wave excitations with $J^P=1^+$ are expected. These two axial $D_1$ states can be identified with the $D_1(2420)\equiv D_1$ and the $D_1(2430)\equiv D_1^\prime$. In the heavy quark limit the two degenerate $1^+$ states are characterized by the angular momentum $j_q$ of the light quark with $j_q=3/2$ and $1/2$. While the strong decay $D_1(j_q=3/2)\to D^\ast \pi$ proceeds by $D$-wave, the transition $D_1(j_q=1/2)\to D^\ast \pi$ has a final $S$-wave. The state decaying via $D$-wave is narrow while the one decaying in an $S$-wave is expected to be broad. Since heavy-light mesons are not charge conjugation eigenstates the axial states can also be written as a superposition of the $^1P_1$ and $^3P_1$ configurations ($J=L$ and $S=0$ or 1) with 
\eq
\begin{aligned}
\big|D_1\big>&=&\cos\phi\big|^1P_1\big>+\sin\phi\big|^3P_1\big>\,,\\
\big|D_1^\prime\big>&=&-\sin\phi\big|^1P_1\big>+\cos\phi\big|^3P_1\big>\,.\label{eq:D1}
\end{aligned}
\en
More detailed analyses~\cite{Barnes:2005pb,vanBeveren:2003jv,Zhong:2008kd} of the mixing scheme in terms of the total width indicate that the mixing angle has a value of about $\phi=\arctan\big(1/\sqrt2\big)\approx35.3^\circ$, the ''magic'' value expected from the heavy quark limit. With this phase convention (alternatively $\phi=-\arctan(\sqrt2)\approx-54.7^\circ$ can be used) the $D_1$ state is identified with the narrow $D_1(2420)$ ($\Gamma\approx20$ MeV~\cite{Amsler:2008zzb}), while the broad $D_1(2430)$ ($\Gamma\approx380$ MeV~\cite{Amsler:2008zzb}) is connected to the $D_1^\prime$. Since our aim is to study the $Z^+$ as a mesonic bound state containing a $D_1$ state the narrow $D_1(2420)$ with its long lifetime is more favorable than the broad $D_1(2430)$. 

Since the $Z(4430)^+$ was observed in the $\psi^\prime \pi^+$ final state, isospin and $G-$parity assignments are $I^G=1^+$. If the $Z(4430)^+$ is a $S$-wave $D_1\overline{D^\ast}$ molecule the $J^P$ quantum numbers are $0^-,\,1^-$ or $2^-$. Here we restrict the study to the $0^-$ and $1^-$ cases since $J^P=2^-$ seems excluded  by the small phase space in the $B\to Z^+K$ production process~\cite{Nielsen:2009uh}.

Following the convention discussed in~\cite{Liu:2007bf,Nielsen:2009uh} the particle content of the isospin multiplet is given as:
\be
\big|Z^+\big>&=&\frac{1}{\sqrt2}\big(\big|D_1^+\overline{D^{\ast\,0}}\big>
+\big|\overline{D^{0}_1}D^{\ast\,+}\big>\big)\,,\nonumber\\
\big|Z^0\big>&=&\frac{1}{2}\big(\big|D_1^+D^{\ast\,-}\big>
-\big|D_1^0\overline{D^{\ast\,0}}\big>+\big|{D_1^-}D^{\ast\,+}\big>-\big|\overline{D_1^0}D^{\ast\,0}\big>
\big)\,,\\
\big|Z^-\big>&=& -\frac{1}{\sqrt2}\, \big(\big|D_1^0D^{\ast\,-}\big>
+\, \big|D^{-}_1D^{\ast^0}\big>\big)\,.\nonumber
\en

In the present method the meson bound state is first set up by the effective interaction Lagrangian between the hadronic molecule and its constituent mesons. In case of $J^P=0^-$ the Lagrangian reads
\eq
{\cal L}_{ZD_1D^\ast}=\frac{g_{_{ZD_1D^\ast}}}{\sqrt2} 
Z^-(x)\int dy\,\Phi(y^2)\, 
\Big\{{D_1^+}^\mu\Big(x-\frac y2\Big){\overline{D^{\ast^0}_\mu}}
\Big(x+\frac y2\Big)+{\overline{D_1^0}}^\mu 
 \Big(x-\frac y2\Big){D^{\ast^+}_\mu}\Big(x+\frac y2\Big)\Big\} 
+ {\rm h.c.}\,. \label{eq:L0}
\en
For $J^P=1^-$ the respective Lagrangian ${\cal L}_{ZD_1D^\ast}$ is given by
\eq
{\cal L}_{ZD_1D^\ast}=i\frac{g_{_{ZD_1D^\ast}}^\prime}{\sqrt2}\epsilon^{\alpha\beta\mu\nu}\partial_\mu Z_\nu^-(x)\int dy\,\Phi(y^2)\Big\{{D_1^+}_\alpha\Big(x-\frac y2\Big){\overline{D^{\ast^0}_\beta}}\Big(x+\frac y2\Big)+{\overline{D_1^0}}_\alpha\Big(x-\frac y2\Big){D^{\ast^+}_\beta}\Big(x+\frac y2\Big)\Big\} 
+ {\rm h.c.} \label{eq:L1}
\en
with $x$ and $y$ being the center of mass and relative coordinates. The correlation function $\Phi(y^2)$ takes into account the distribution of the constituent mesons in the $Z$ resonance. Its Fourier transform enters as a form factor in our analysis which also leads to a regularization of the structure integrals. In earlier works~\cite{Anikin:1993mm,Anikin:1995cf} it was found that observables as decay widths which are studied in this paper are not sensitive to the specific shape of this form factor as long as the intrinsic scale remains the same. In the present case we deal with the Gaussian function 
\eq
\Phi(y^2)=\int\frac{d^4k}{(2\pi)^4}e^{-iky}\widetilde \Phi(-k^2),\quad\widetilde\Phi(k_E^2)=\exp(-k_E^2/\Lambda^2)\,,
\en
where the index $E$ refers to the Euclidean momentum. The size parameter $\Lambda$ is fixed in the physically meaningful region of a few GeV. In the present work we study finite size effects by varying $\Lambda_Z$ in the range of  $1.5-2.5$ GeV. In the special case of a pointlike molecular structure the correlation function $\Phi(y^2)$ reduces to the delta function $\delta^4(y)$, equivalent to $\widetilde\Phi(k_E^2)\rightarrow 1$.
\begin{figure}[b]
\includegraphics[scale=0.6]{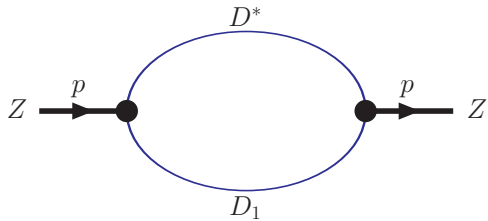}
\caption{Mass operator of the $Z(4430)$ meson}
\label{fig1}
\end{figure}

The coupling of $Z$ to the virtual constituents, denoted by $g_{ZD_1D^\ast}$, is fixed by means of the compositeness condition~\cite{Weinberg:1962hj,Salam:1962ap,Efimov:1993zg,Branz:2008ha,Ivanov:1996pz,Ivanov:1996fj,%
Branz:2009yt,Faessler:2007gv,Faessler:2007us,Faessler:2007cu,Dong:2008gb,%
Dong:2009yp,Dong:2009tg,Dong:2009uf,Dong:2010gu} 
which provides a self-consistent method to adjust this quantity. Since we deal with a hadronic bound state, the physical field of the $Z$ meson should be fully expressed by the fields of the constituent mesons and therefore does not contain a bare component. This is achieved by setting the field renormalization constant $Z_{Z}$ to zero with
\eq
Z_{Z}=1-g_{_{ZD_1D^\ast}}^{2}\tilde\Sigma^\prime (m_{Z}^2)=0\,,\label{eq:cc}
\en
where $\tilde\Pi^\prime(m_{Z}^2)=g_{_{ZD_1D^\ast}}^2\tilde\Sigma^\prime(m_{Z})$ is the derivative of the mass operator illustrated by the diagram in Fig.~\ref{fig1}. 
If the $Z$ meson is a vector-like object (quantum numbers $J^P=1^-$) the mass operator is split into its transverse and longitudinal parts $\Sigma$ and $\Sigma^L$
\eq
\Sigma^{\mu\nu}(p^2)=\Sigma(p^2)g_\perp^{\mu\nu}+\Sigma^L(p^2)\frac{p^\mu p^\nu}{p^2}\,,
\en
where $g_\perp^{\mu\nu}=g^{\mu\nu}-\frac{p^\mu p^\nu}{p^2}$ and $g^{\mu\nu}_\perp p_\mu=0$. In the compositeness condition of Eq. (\ref{eq:cc}) only the transverse part enters. 

Meson loop diagrams are evaluated by using the free meson propagators, 
which in momentum space read as 
\eq
\tilde S_{H_1}(k) &=& \frac{1}{M_{H_1}^2 - k^2 - i\epsilon}
\en
in case of pseudoscalar and scalar mesons ($H_1=P,S$). For the case of vector and axial-vector mesons ($H_2=V,A$) we use
\eq
\tilde S_{H_2}^{\mu\nu}(k) &=& 
\frac{-g^{\mu\nu}+k^\mu k^\nu/M_{H_2}^2}{M_{H_2}^2 - k^2 - i\epsilon}\,. 
\label{eq:prop}
\en 
Here we do not include the finite widths of the constituent mesons in the propagators (especially for the $D_1(2420)$ with a total width of about 20 MeV) since this effect is negligible in the present analysis. 

The mass values for the intermediate and final state mesons are taken 
from~\cite{Amsler:2008zzb}. 
For convenience we also introduce the binding energy $\epsilon$ defined by the difference between the central value of the $Z^+$ mass and the lower threshold ($D_1^+\overline{D^{\ast\,0}}$) with 
\eq
m_{Z^+}=m_{D_1^+}+m_{\overline D^{\ast\,0}}-\epsilon\,.
\en
We finally present our results in dependence on the possible values of the binding energy $\epsilon$.

\section{radiative decay}\label{sec:radiative}

\begin{figure}[thbp]
\includegraphics[scale=0.6]{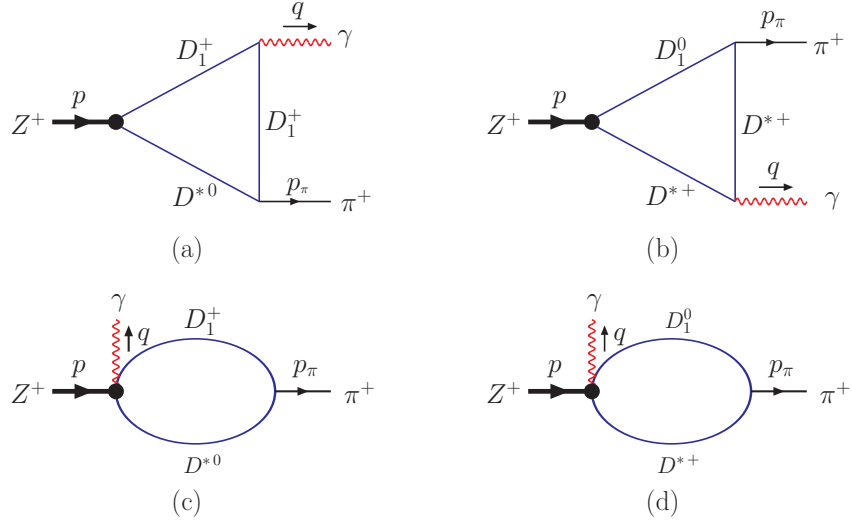}
\caption{Diagrams contributing to the radiative $Z^+\to \pi^+\gamma$ decay.}
\label{fig2}
\end{figure}

We first consider the radiative decay $Z^+\to \pi^+\gamma$ which in the molecular $D_1\overline{D^\ast}$ interpretation proceeds by the diagrams of Fig. \ref{fig2}. In the following we discuss the relevant vertices entering in the radiative decay process. 

One of the interaction Lagrangians relevant  for the radiative decay arises from gauging the free Lagrangians by using minimal substitution
\eq
\partial^\mu M^\pm\rightarrow (\partial^\mu\mp ieA^\mu)M^\pm\,,
\en
which leads to 
\eq
{\cal L}^{em(1)}=ieA_\mu\big(Z^-\partial^{\mu^{^{\!\!\!\!\!\!\!\!\leftrightarrow}}}Z^++\pi^-\partial^{\mu^{^{\!\!\!\!\!\!\!\!\leftrightarrow}}}\pi^++g^{\alpha\beta}V^{-}_\alpha\partial^{\mu^{^{\!\!\!\!\!\!\!\!\leftrightarrow}}}V^{+}_\beta+g^{\mu\beta}V^-_\alpha\partial^\alpha V^+_\beta-g^{\mu\alpha}\partial^\beta V^-_\alpha V^+_\beta\big)\,,
\en
where 
$A\partial^{\mu^{^{\!\!\!\!\!\!\!\!\leftrightarrow}}}B=
A\partial^\mu B-B\partial^\mu A$ and $V=D^\ast,\,D_1$. The corresponding vertices enter in Figs. \ref{fig2}~(a) and \ref{fig2}~(b), where the triangle diagrams are obtained by coupling the final states to the $Z^+$ constituents. These graphs yield the dominant contributions to the decay amplitude. Because of their nonlocal structure the strong interaction Lagrangians (\ref{eq:L0}) and (\ref{eq:L1}) are not invariant under $U_{em}(1)$ transformations and need to be modified accordingly. We use the method suggested in~\cite{Terning:1991yt} where each charged meson field $M$ is multiplied by an exponential containing the gauge field
\eq
M^\pm(y)\rightarrow e^{\mp iI(y,x,P)}M^\pm(y)
\en
and $I(y,x,P)=\int\limits_x^y\,dz_\mu A^\mu(z)$. This modification leads to further interaction vertices contained in Figs. \ref{fig2}~(c) and \ref{fig2}~(d). These additional graphs \ref{fig2}~(c) and \ref{fig2}~(d) are strongly suppressed but they have to be included to guarantee full gauge invariance. 

The interaction between the final pion and the charmed mesons $D_1$, $D^\ast$ in the loops is set up by the interaction Lagrangian
\eq
{\cal L}_{D_1D^\ast\pi}&=&\frac{g_{_{D_1D^\ast\pi}}}{2\sqrt2}\,
D_1^{\mu\nu} \, {\bm\pi} \, {\bm\tau} \, \overline{D^\ast}_{\mu\nu} + {\rm h.c.} 
\,,\label{eq:L4}
\en 
where $V^{\mu\nu}=\partial^\mu V^\nu-\partial^\nu V^\mu$ is the stress tensor of the vector mesons $V=\psi,D_1$ and $D^\ast$. The interaction vertex involving the $D_1(2420)$ meson should contain a dominant $D$-wave $D^\ast\pi$ coupling (see e.g.~\cite{Barnes:2005pb,vanBeveren:2003jv,Godfrey:1997ek}) which dictates the form of the Lagrangian in  (\ref{eq:L4}) with two derivatives involved. This dynamical selection rule, obtained in the heavy quark limit, also leads to the form of the Lagrangian (\ref{eq:L3}) in case of the strong decays discussed in the next section. The coupling constant $g_{D_1D^\ast\pi}$ is derived from the width of $D_1\to D^{\ast} \pi$ which is the dominant decay mode of $D_1$. The partial decay width is expected to be around 20~MeV~\cite{Zhong:2008kd}, where $\Gamma(D_1^0\to D^{\ast\,+} \pi^-)=2\, \Gamma(D_1^0\to D^{\ast\,0} \pi^0)$. The decay width is set up as  
\eq
\Gamma(D_1^0\to D^{\ast\,+} \pi^-)&=&\frac{\lambda^{1/2}(m_{D_1}^2,m_{D^\ast}^2,m_\pi^2)}{16\pi\,m_{D_1}^3}\overline{\big|{\cal M}\big|^2}\,,
\en
where $\lambda(a,b,c)=a^2+b^2+c^2-2ab-2ac-2bc$ denotes the K\"allen function and 
$\overline{\big|{\cal M}\big|^2} = 1/3 \sum\limits_{\rm pol} \big|{\cal M}\big|^2$ 
represents the spin-averaged and summed over polarizations transition amplitude squared. The effective $D$-wave interaction Lagrangian of Eq. (\ref{eq:L4}) leads to the matrix element
\eq
{\cal M}_{D_1^0\to D^{\ast\,+} \pi^-}^{\mu\nu}=g_{D_1D^\ast\pi}\big(p_{D^\ast}p_{D_1}g^{\mu\nu}-p_{D^\ast}^\mu p_{D_1}^\nu\big)\,.
\en
Hence, the decay width for a $D$-wave decay is of the form
\eq
\Gamma(D_1^0\to D^{\ast\,+} \pi^-)&=&\frac{g_{D_1D^\ast \pi}^2}{96\pi m_{D_1}^3}\lambda^{1/2}\big(\lambda+6m_{D_1}^2m_{D^\ast}^2\big)
\en
with the resulting coupling $g_{D_1D^\ast\pi}=0.49$ GeV$^{-1}$ for $\Gamma(D_1\to D^{\ast}\pi)\approx$20 MeV.

In case of $J^P=0^-$ the radiative decay 
$Z^+\to\pi^+\gamma$ is excluded due to gauge invariance. 
If the spin-parity of the $Z^+$ is $1^-$ then  
the transition amplitude ${\cal M}^{\mu\nu}$ has 
the structure 
\eq
{\cal M}^{\mu\nu} = e \, F_{Z\pi\gamma}(m_Z^2,m_\pi^2,0)\, 
\epsilon^{\alpha\beta\mu\nu}p_\alpha q_\beta\,, 
\en 
where $g_{Z\pi\gamma}$ is the effective coupling. It is related to the corresponding transition form factor 
$g_{Z\pi\gamma}\equiv F_{Z\pi\gamma}(m_Z^2,m_\pi^2,0)$  
evaluated via the loops of Fig. \ref{fig2}. Finally, in terms of the effective coupling $g_{Z\pi\gamma}$ 
the decay width $\Gamma(Z^+\to\pi^+\gamma)$ is given by
\eq
\Gamma(Z^+\to\pi^+\gamma) = 
\frac{\alpha}{24} \, 
g_{Z\pi\gamma}^2 \, m_Z^3 \, 
\Big( 1 - \frac{m_\pi^2}{m_Z^2} \Big)^3 \simeq 
\frac{\alpha}{24} \,  
g_{Z\pi\gamma}^2 \, m_Z^3 \,. 
\en

\section{strong hidden-charm decays}\label{strong}

\begin{figure}[thbp]
\includegraphics[scale=0.6]{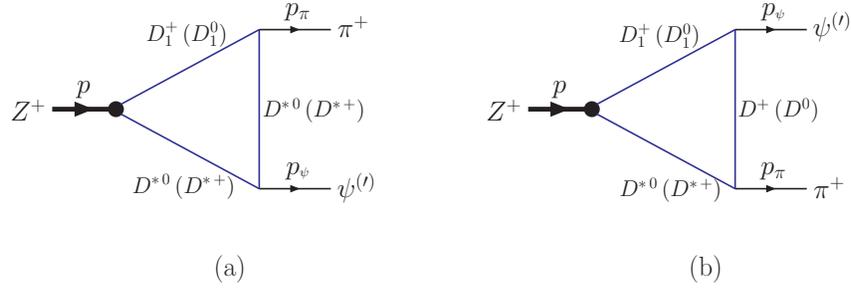}
\caption{Diagrams contributing to the $Z^+\to \psi^{(\prime)}\pi^+$ decay.}
\label{fig3}
\end{figure}

In order to study the strong hidden-charm decays $Z^+\to \psi^{(\prime)}\pi^+$ we first set up the interaction between the final state and the constituent $D_1$ and $D^\ast$ mesons. We use the effective Lagrangians
\eq
{\cal L}_{D^\ast D\pi}&=&\frac{g_{_{D^\ast D\pi}}}{2\sqrt2}\,D^{\ast\,\dagger}_\mu\, {\bm\pi} \, {\bm\tau} 
\, i\partial^{\mu^{^{\!\!\!\!\!\!\!\!\leftrightarrow}}}\, D + {\rm h.c.}\,,\label{eq:L2}\\
{\cal L}_{D^\ast D^\ast\psi}&=&ig_{_{D^\ast D^\ast\psi}}\,\big(\psi^{\mu\nu}\overline{D^\ast_\mu} D^\ast_\nu+\psi^\mu\overline{D^{\ast}}^\nu D^\ast_{\mu\nu}+\psi^\nu\overline{D^\ast}_{\mu\nu}{D^\ast}^\mu\big)\,,\label{eq:L5}\\
{\cal L}_{D_1D\psi}&=&\frac{g_{_{D_1D\psi}}}{2}\,D_1^{\mu\nu}\psi_{\mu\nu}D + {\rm h.c.}\,.\label{eq:L3}
\en
The coupling of the $Z^+$ to its constituents is determined by Eq. (\ref{eq:cc}). In addition, ${\cal L}_{D_1D^\ast\pi}$ was already defined in (\ref{eq:L4}) in the framework of radiative decays. The coupling strengths $g_{D^\ast D\pi}=17.9$ and $g_{D^\ast D^\ast J/\psi}\approx8$ as well as the ratio $g_{\psi^\prime D^\ast D^\ast}/g_{\psi D^\ast D^\ast}$ are taken from heavy hadron chiral perturbation theory (HHChPT)~\cite{Dong:2009uf,Faessler:2007gv}. This ratio of couplings of excited $\psi^\prime$ to $\psi$ is fixed by $g_{\psi^\prime D^\ast D^\ast}/g_{\psi D^\ast D^\ast}=m_{\psi^\prime}f_{\psi}/(m_\psi f_{\psi^\prime})=1.67$, where $f_{\psi^{(\prime)}}$ is the leptonic decay constant. An estimate for the remaining coupling $g_{D_1D\psi}$ is taken from a coupled channel analysis (see e.g.~\cite{Gamermann:2007fi}). We use $|g_{D_1D^\ast\pi}|\approx72$ MeV and $|g_{D_1D\psi}|\approx29$ MeV from~\cite{gamermann} in order to estimate the ratio of the coupling strengths $r_1=g_{D_1D\psi}/g_{D_1D^\ast\pi}\approx0.4\pm0.2$, where we assumed an uncertainty of 50\%. By using the ratio $r_1$ and $g_{D_1D\pi}=0.49$ GeV$^{-1}$ as defined above, we can give a rough estimate of $g_{D_1D\psi}\approx0.2\pm0.1$ GeV$^{-1}$. As discussed below, the diagram (a) of Fig. \ref{fig3} clearly dominates the hidden-charm decay.  Therefore, in leading order the $Z^+\to \psi^{(\prime)} \pi^+$ transition can be regarded to be proportional to $g_{D_1D\psi^{(\prime)}}$. A variation of $g_{D_1D\psi^{(\prime)}}$ modifies the decay width accordingly. In order to estimate the ratio of couplings $r_2=g_{D_1D\psi^\prime}/g_{D_1D\psi}$ we use the $^3P_0$ model~\cite{LeYaouanc:1972ae,*LeYaouanc:1973xz}, where the details of this procedure can be found in Appendix \ref{A:3P0}. We find for the ratio of couplings $g_{D_1D\psi^\prime}/g_{D_1D\psi}$ a value which is close to 2 and of the same order as the above mentioned ratio $g_{\psi^\prime D^\ast D^\ast}/g_{\psi D^\ast D^\ast}=1.67$. This hierarchy of couplings involving the $\psi$ and $\psi^{\prime}$ charmonium states is consistent with the HHChPT scaling indicated above. Further on we include uncertainties in the predictions of the $^3P_0$ model which for example might arise due to variations of the quark pair production amplitude (as e.g. discussed in~\cite{Downum:2006re}) which is usually fitted to data. In literature the quark-pair production strength ranges between 0.4~\cite{Barnes:2005pb} to 0.5~\cite{Ackleh:1996yt}. We therefore consider an uncertainty of 50\% in the ratio $r_2=g_{D_1D\psi^\prime}/g_{D_1D\psi}=2\pm1$.

The diagrams describing the hidden charm decay are illustrated in 
Fig. \ref{fig3}.

Provided that the $Z^+$ is a pseudoscalar the transition amplitude for the hidden charm decay mode can be expressed by two form factors $F_{1,2}$:
\eq
{\cal M}^\mu_{Z^+\to\pi^+\psi}=F_1(m_Z^2,m_\psi^2,m_\pi^2)p_\pi^\mu+F_2(m_Z^2,m_\psi^2,m_\pi^2)p_\psi^\mu\,.
\en
Here only the first form factor contributes to the decay width 
$\Gamma(Z^+\to \psi^{(\prime)}\pi^+)$: 
\eq
\Gamma(Z^+\to \psi^{(\prime)}\pi^+)=\frac{\lambda^{1/2}}{16\pi m_{Z}^3}\overline{\big|{\cal M}\big|^2}=g_{_{Z\pi\psi}}^2\frac{\lambda^{\frac32}(m_Z^2,m_\psi^2,m_\pi^2)}{64\pi m_Z^3m_\psi^2}\,, 
\en 
where $g_{_{Z\pi\psi}}\equiv F_1(m_Z^2,m_\psi^2,m_\pi^2)$ and 
$\lambda(x,y,z)=x^2+y^2+z^2-2xy-2xz-2yz$ is the K\"allen function.

If we deal with a vector $Z^+$ the matrix element is given by
\eq
{\cal M}^{\mu}&=&F_3(m_Z^2,m_\psi^2,m_\pi^2)\epsilon^{\alpha\beta\mu\nu}p_{\pi\,\alpha}p_\beta
\en
and by analogy we use $g_{Z\pi\psi}^\prime \equiv F_3(m_Z^2,m_\psi^2,m_\pi^2)$ in order to calculate the decay width
\eq
\Gamma(Z^+\to \psi^{(\prime)}\pi^+)=g^{\prime\,2}_{_{Z\pi\psi}}\frac{\lambda^{\frac32}(m_Z^2,m_\psi^2,m_\pi^2)}{96\pi m_Z^3}\,.
\en

%%%%%%%%%%%%%%%%%%%%%%%%%%%
\section{Results}\label{sec:results}
\def\arraystretch{1.5} 

In the following section we discuss our results on the decay properties of the $Z^+$. As far as the hadronic molecule is involved the coupling constant is fixed by the compositeness condition. The corresponding couplings are summarized in Table  \ref{tab0} for the quantum numbers $J^P=0^-$ and $1^-$. Values for the coupling constants are presented in dependence on the binding energy $\epsilon$. Since the $Z^+$ mass lies close to the ($D_1^+\overline{D^{\ast\,0}}$) threshold at 4.43 GeV the values for $\epsilon$ are of a few MeV, in particular we vary $\epsilon$ between 1 and 10 MeV. The errors on the numerical results are due to variations of the model parameter $\Lambda_Z$ from 1.5 to 2.5 GeV.
\begin{table}[htbp]
\caption{Coupling constants $g_{ZD_1D^\ast}$ ($J^{P}=0^-$) and $g_{ZD_1D^\ast}^\prime$ ($J^{P}=1^-$) in GeV  for $\Lambda_Z=1.5-2.5$ GeV and $\epsilon$=$1-10$ MeV.}
\label{tab0}
\renewcommand{\arraystretch}{1.4}
     \setlength{\tabcolsep}{0.5cm}
     \centering
\begin{tabular}{ccccc}
\hline\hline
$\epsilon$ [MeV]&1&5 &10\\\hline
$g_{ZD_1D^\ast}$ ($J^{P}=0^-$) &$3.8\pm0.1$&$5.6\pm0.1$&$6.8\pm0.2$\\\hline
$g_{ZD_1D^\ast}^\prime$ ($J^{P}=1^-$) &$1.2\pm0.1$&$1.8\pm0.1$&$2.1\pm0.1$\\\hline
\hline
\end{tabular}

\vspace*{.2cm}

\caption{Decay widths $\Gamma_{Z^+\to\pi^+\gamma}$ in keV for $J^{P}=1^-$ with $\epsilon=1-10$ MeV and $\Lambda_Z=1.5-2.5$ GeV.}
\label{tab3}
\renewcommand{\arraystretch}{1.4}
     \setlength{\tabcolsep}{0.4cm}
     \centering
\begin{tabular}{cccc}
\hline\hline
$\epsilon$ [MeV]& 1&5&10\\\hline\hline
$\Gamma_{Z^+\to\pi^+\gamma}$ [keV]&$0.3^{+0.2}_{-0.1}$&$0.6^{+0.3}_{-0.3}$&$0.8^{+0.4}_{-0.3}$\\
\hline\hline
\end{tabular}
\end{table}

\begin{table}[hbp]
\caption{Decay widths $\Gamma_{Z\to\psi^{(\prime)}\pi}$ in MeV for $J^P=0^-$, $\Lambda_Z=1.5-2.5$ GeV and $r_2=g_{D_1D\psi^\prime}/g_{D_1D\psi}=1-3$.}
\label{tab4}
\renewcommand{\arraystretch}{1.4}
     \setlength{\tabcolsep}{0.2cm}
     \centering
\begin{tabular}{l|cccccccc}
\hline\hline
&$\epsilon$ [MeV]&$\Gamma_{Z\to\pi\psi}$ [MeV]&\multicolumn{3}{c}{$\Gamma_{Z\to\pi\psi^\prime}$ [MeV]}&\multicolumn{3}{c}{$R=\Gamma_{Z\to\pi\psi^\prime}/\Gamma_{Z\to\pi\psi}$}\\
&&&$r_2=1$&$r_2=2$&$r_2=3$&$r_2=1$&$r_2=2$&$r_2=3$\\\hline\hline
\multirow{3}{*}{$\frac{g_{D_1D\psi}}{g_{D_1D^\ast \pi}}=0.2$}&1&$0.2^{+0.1}_{-0.1}$&$0.1^{+0.0}_{-0.1}$&$0.3^{+0.1}_{-0.1}$&$0.8^{+0.2}_{-0.3}$&$\approx0.2$&$\approx1.5$&$\approx3.8$\\\cline{2-9}
&5&$0.4^{+0.2}_{-0.2}$&$0.1^{+0.1}_{-0.0}$&$0.6\pm0.2$&$1.5^{+0.4}_{-0.5}$&$\approx0.3$&$\approx1.5$&$\approx3.9$\\\cline{2-9}
&10&$0.5^{+0.3}_{-0.2}$&$0.1^{+0.1}_{-0.0}$&$0.8^{+0.2}_{-0.3}$&$2.0^{+0.5}_{-0.6}$&$\approx0.3$&$\approx1.5$&$\approx3.7$\\\hline\hline
\multirow{3}{*}{$\frac{g_{D_1D\psi}}{g_{D_1D^\ast \pi}}=0.4$}&1&$0.9^{+0.3}_{-0.3}$&$0.3^{+0.1}_{-0.1}$&$1.5^{+0.4}_{-0.4}$&$3.5^{+0.9}_{-0.9}$&$\approx0.3$&$\approx1.6$&$\approx4.0$\\\cline{2-9}
&5&$1.7^{+0.5}_{-0.6}$&$0.6^{+0.2}_{-0.2}$&$2.8^{+0.6}_{-0.8}$&$6.7^{+1.3}_{-1.7}$&$\approx0.3$&$\approx1.7$&$\approx4.0$\\\cline{2-9}
&10&$2.3^{+0.7}_{-0.8}$&$0.8^{+0.2}_{-0.3}$&$3.7^{+0.8}_{-1.1}$&$8.7^{+1.8}_{-2.3}$&$\approx0.3$&$\approx1.6$&$\approx3.8$\\
\hline\hline
\multirow{3}{*}{$\frac{g_{D_1D\psi}}{g_{D_1D^\ast \pi}}=0.6$}&1&$2.1^{+0.5}_{-0.7}$&$0.8^{+0.2}_{-0.3}$&$3.5^{+0.9}_{-0.9}$&$8.3^{+1.9}_{-2.0}$&$\approx0.4$&$\approx1.7$&$\approx4.0$\\\cline{2-9}
&5&$3.9^{+1.2}_{-1.2}$&$1.5^{+0.4}_{-0.5}$&$6.7^{+1.3}_{-1.7}$&$15.6^{+2.8}_{-3.8}$&$\approx0.4$&$\approx1.7$&$\approx4.0$\\\cline{2-9}
&10&$5.3^{+1.5}_{-1.7}$&$2.0^{+0.5}_{-0.6}$&$8.7^{+1.8}_{-2.3}$&$20.1^{+3.9}_{-4.9}$&$\approx0.4$&$\approx1.6$&$\approx3.8$\\
\hline\hline
\end{tabular}

\vspace*{.2cm} 

\caption{Decay widths $\Gamma_{Z\to\psi^{(\prime)}\pi}$ in MeV for $J^P=1^-$, $\Lambda_Z=1.5-2.5$ GeV and $r_2=g_{D_1D\psi^\prime}/g_{D_1D\psi}=1-3$.}
\label{tab5}
\renewcommand{\arraystretch}{1.4}
     \setlength{\tabcolsep}{0.2cm}
     \centering
\begin{tabular}{l|cccccccc}
\hline\hline
&$\epsilon$ [MeV]&$\Gamma_{Z\to\pi\psi}$ [MeV]&\multicolumn{3}{c}{$\Gamma_{Z\to\pi\psi^\prime}$ [MeV]}&\multicolumn{3}{c}{$R=\Gamma_{Z\to\pi\psi^\prime}/\Gamma_{Z\to\pi\psi}$}\\
&&&$r_2=1$&$r_2=2$&$r_2=3$&$r_2=1$&$r_2=2$&$r_2=3$\\\hline\hline
\multirow{3}{*}{$\frac{g_{D_1D\psi}}{g_{D_1D^\ast \pi}}=0.2$}
&1&$0.1$&0.1&$0.3^{+0.1}_{-0.1}$&$0.8^{+0.2}_{-0.1}$&$\approx0.8$&$\approx5.5$&$\approx14.0$\\\cline{2-9}
&5&$0.1$&$0.1$&$0.6^{+0.0}_{-0.1}$&$1.5^{+0.2}_{-0.3}$&$\approx0.8$&$\approx5.3$&$\approx13.6$\\\cline{2-9}
&10&$0.1$&$0.1$&$0.7^{+0.1}_{-0.1}$&$1.9^{+0.3}_{-0.4}$&$\approx0.8$&$\approx5.2$&$\approx13.6$\\\hline\hline
\multirow{3}{*}{$\frac{g_{D_1D\psi}}{g_{D_1D^\ast \pi}}=0.4$}&1&$0.4^{+0.1}_{-0.1}$&$0.3-0.4$&$1.6^{+0.3}_{-0.4}$&$3.8^{+0.7}_{-0.8}$&$\approx0.7$&$\approx3.5$&$\approx8.3$\\\cline{2-9}
&5&$0.8^{+0.2}_{-0.2}$&$0.6^{+0.1}_{-0.1}$&$2.9^{+0.4}_{-0.7}$&$6.9^{+1.0}_{-1.6}$&$\approx0.8$&$\approx3.6$&$\approx8.6$\\\cline{2-9}
&10&$1.1^{+0.2}_{-0.3}$&$0.7_{-0.1}^{+0.1}$&$3.6^{+0.6}_{-0.7}$&$8.7^{+1.5}_{-1.9}$&$\approx0.6$&$\approx3.3$&$\approx7.9$\\\hline\hline
\multirow{3}{*}{$\frac{g_{D_1D\psi}}{g_{D_1D^\ast \pi}}=0.6$}&1&$1.2^{+0.3}_{-0.3}$&$0.8^{+0.2}_{-0.1}$&$3.8^{+0.7}_{-0.8}$&$9.0^{+1.7}_{-2.1}$&$\approx0.7$&$\approx3.1$&$\approx7.4$\\\cline{2-9}
&5&$2.3^{+0.4}_{-0.6}$&$1.5^{+0.2}_{-0.3}$&$6.9^{+1.0}_{-1.6}$&$16.2^{+2.6}_{-3.8}$&$\approx0.7$&$\approx3.0$&$\approx7.0$\\\cline{2-9}
&10&$2.9^{+0.6}_{-0.8}$&$1.9^{+0.3}_{-0.4}$&$8.7^{+1.5}_{-1.9}$&$20.4^{+3.6}_{-4.5}$&$\approx0.7$&$\approx3.0$&$\approx7.0$\\\hline\hline
\end{tabular}
\end{table}

The radiative decay width for $Z^+\to \pi^+\gamma$ is analyzed for $1^-$, whereas for $J^P=0^-$ it is forbidden. The results are indicated in Tab. \ref{tab3} for different values of the binding energy and the size parameter. Our predictions are in general rather sizable and of the order of 0.2 to 1.2 keV. In Tab. \ref{tab3} the smaller value of each entry corresponds to $\Lambda_Z=1.5$~GeV while the larger one is related to $\Lambda_Z=2.5$ GeV.

The decay widths for the hidden charm decay channels $Z\to\pi\psi$ and $Z\to \pi\psi^\prime$ are given in Tab. \ref{tab4} for $J^P=0^-$ and in Tab. \ref{tab5} in case of $1^-$. In both tables the error bars indicate changes in the finite size effects with $\Lambda_Z$ varied from 1.5 to 2.5 GeV. The ratios $r_1=\frac{g_{D_1D\psi}}{g_{D_1D^\ast\pi}}=0.4\pm0.2$ and  $r_2=\frac{g_{D_1D\psi^\prime}}{g_{D_1D\psi}}=2\pm1$ we fixed from values obtained in coupled channel analyses~\cite{Gamermann:2007fi} and from the phenomenology of the $^3P_0$ model (see Appendix). We consider uncertainties in the ratios $r_1$ and $r_2$ and study the decay properties of the $Z^+$ for possible values of the coupling constants.

The hidden charm decay mode is generated by the graphs of Fig.~\ref{fig3}, where diagram \ref{fig3}~(b) clearly dominates the transition amplitude by one order of magnitude in comparison to graph \ref{fig3}~(a). This dominance results from the sizable $D^\ast D\pi$-coupling but also from the lighter $D$-meson mass (compared to $D^\ast$ exchange) in the rescattering process. As a consequence the decay widths are very sensitive to variations of the couplings  $g_{D_1D\psi^{(\prime)}}$; variations of $r_1$ and $r_2$ enter approximately quadratically in the decay widths. For instance, $\Gamma_{Z\to\pi\psi^\prime}/\Gamma_{Z\to\pi\psi}\propto r_2^2$ as can be read off from the results in the last three columns of Tabs. \ref{tab4} and \ref{tab5}.

For values of the ratios in the region $r_1= g_{D_1D\psi}/g_{D_1D^\ast\pi}=0.4$ to 0.6 and $r_2= g_{D_1D\psi^\prime }/g_{D_1D\psi}=2-3$ the decay width $\Gamma_{Z\to\pi\psi^\prime}$ is in the MeV range which is consistent with the expectation from observation~\cite{Godfrey:2008nc}. However, if the ratio $r_1$ is relatively small as in the case of $r_1=0.2$ the decay width $\Gamma_{Z\to\pi\psi^\prime}$ becomes smaller than one MeV, which seems excluded by observation.  

For the quantum numbers $J^P=0^-$ the decay mode $Z\to  \pi\psi$ is suppressed relative to $\pi\psi^\prime$ by a factor $\approx2$ for $r_2=2$ and about $4$ for $r_2=3$, respectively. In case of $1^-$  the ratio of decay rates $R=\Gamma(Z^+\to\pi^+\psi^\prime)/\Gamma(Z^+\to\pi^+\psi)$ is even larger with $R\approx3$ for $r_2=2$ and $R\approx8$ for $r_2=3$ which is at least qualitatively in line with the experimental observation that the $\pi^+\psi^\prime $ decay mode dominates the $ \pi^+\psi$ partial decay width.  The ratio $R$ is rather insensitive to variations of the model parameter $\Lambda_Z$ and the binding energy $\epsilon$ as indicated in Tabs. \ref{tab4} and  \ref{tab5} for $J^P=0^-$ and $J^P=1^-$. On the contrary, for equal couplings $g_{D_1D\psi^\prime}$ and $g_{D_1D\psi}$, i.e. $r_2=1$, the branching ratio $R=\Gamma(Z^+\to\pi^+\psi^\prime)/\Gamma(Z^+\to\pi^+\psi)$ is smaller than one which presently is in contradiction to experimental observations.

At least in the molecular scenario, the radiative decay might help to partially settle the $J^P$ quantum numbers of the $Z$, the strong $Z^+\to  \pi^+\psi^\prime$ decay is not very sensitive to the choice of $J^P$. The dependence on $J^P$ of the strong hidden-charm decay widths is not very pronounced, for $J^P=1^-$ decay widths are only slightly smaller than for $0^-$. 

Within  the hadronic bound state interpretation hidden-charm decays were  also discussed in~\cite{Meng:2007fu}. But in the present evaluation  the contributions of the $D^\ast$ and $D$ rescattering processes represented in Fig.~\ref{fig3} are significantly different. In our case diagram \ref{fig3}~(b) is dominant because of the sizable coupling of the $D^\ast$ meson to the $D\pi$ mode. On the contrary in~\cite{Meng:2007fu} the diagram in Fig.~\ref{fig3}~(b) is highly suppressed by a factor of about $15-30$ compared to the $D^\ast$ exchange process, which was explained by the small $g_{D_1D\psi}$ coupling. 
We want to remind that the prediction for $R=\Gamma(Z^+\to \pi^+\psi^\prime)/\Gamma(Z^+\to \pi^+\psi)$ primarily depends on the explicit values of the coupling $g_{D_1D\psi^{(\prime)}}$ and therefore on the ratio $g_{D_1D\psi^\prime}/g_{D_1D\psi}$, while variations of $\Lambda_Z$ and $\epsilon$ only play a minor role. In comparison, in~\cite{Meng:2007fu} sizable values for the ratio $R$ are only obtained for small hidden charm decay widths.

\section{conclusions}\label{sec:conclusions}

In the present work we consider hidden charm and radiative decays of the $Z(4430)^+$ in a $D_1(2420)\overline{D^\ast}+h.c.$ molecular structure interpretation. 
As guided by previous studies of possible binding mechanisms in this system we choose the preferred $J^P=0^-$ and $1^-$ quantum numbers for the $Z^+$.

In the predictions for the decay widths we study the influence of finite size effects and the dependence on the exact value of the binding energy.
We give a first prediction for the radiative decay width $Z(4430)^+\to \pi^+\gamma$ only allowed for $J^P=1^-$, which is about 0.5 to 1~keV. This order keV result should allow for a possible detection.
We also analyzed the hidden charm decays $Z^+\to\pi^+\psi$, presently not observed, and $Z^+\to\pi^+\psi^\prime$, the discovery mode of the $Z^+$. Both decays are generated by $D^\ast$ or $D$ rescattering processes. 
Predictions depend crucially on explicit values for the couplings $g_{D_1D\psi^{(\prime)}}$. For approximate values of these couplings we took guidance from a coupled channel analysis~\cite{Gamermann:2007fi,gamermann} (which essentially fixes $g_{D_1D\psi}$) and in addition from the $^3P_0$ model (for the ratio $r_2=g_{D_1D\psi^\prime}/g_{D_1D\psi}$).
Low values for $r_2\approx 1$ generate predictions for $\Gamma_{Z\to\pi\psi^\prime}$ which are much to small to justify observation and for $R=\Gamma_{Z\to \pi\psi^\prime}/\Gamma_{Z\to \pi\psi}$ which contradicts the nonobservation of the $\pi\psi$ mode. Intermediate and larger values for $r_2\approx2-3$ lead to order MeV predictions for $\Gamma_{Z\to\pi\psi^\prime}$ and also to a ratio $R\approx2-9$ at least qualitatively in line with experimental constraints. 
For $J^P=1^-$ the ratio $R$ is slightly larger, but in general a sizable dependence of the predictions on the choice of $J^P$ is not observed. Also, finite size effects (as contained in the size parameter $\Lambda_Z$) and the exact value of the binding energy do not have a large influence on the predictions.

Present predictions have a slight tendency to support the $D_1\overline{D^\ast}$ bound state interpretation of the $Z(4430)^+$ at least what concerns the hidden charm decay modes. A further evaluation of open charm decay modes does not seem to be decisive since predictions also depend crucially on principally unknown coupling constants~\cite{Liu:2008yy}. 
In this respect the radiative mode $\pi^+\gamma$ offers a further test for the structure issue of the $Z^+$. 

From the experimental side it is clear that the existence and, if possible, quantum numbers of the $Z(4430)^+$ still have to be firmly established.
As long as the charged charmonium-like structures are not ruled out they remain an interesting object for future experiments e.g. the $Z^+$ could be studied in nucleon-antinucleon scattering processes in the upcoming PANDA experiment~\cite{Ke:2008kf}.

\begin{acknowledgments}
This work was supported by the DFG under Contract No. FA67/31-2 and No. GRK683. This research is also part of the European
Community-Research Infrastructure Integrating Activity
``Study of Strongly Interacting Matter'' (acronym HadronPhysics2,
Grant Agreement No. 227431), Russian President grant
``Scientific Schools''  No. 3400.2010.2, Russian Science and
Innovations Federal Agency contract No. 02.740.11.0238.
\end{acknowledgments}

\newpage
%%%%%%%%%%%%%%%%%%%%%%%%%%%%BIBLIOGRAPHY%%%%%%%%%%%%%%%%%%%%%%%%%%%%%%%%%%%%%%%

%\begin{thebibliography}{99}

%\end{thebibliography}
%\bibliography{nc,cds,sw,connes}
\bibliography{literature}
%\end{document}
%%%%%%%APPENDIX%%%%%%%%%%%%
\newpage
\begin{appendix}
\section{Ratio of couplings $\mathbf{g_{D_1D\psi^\prime}/g_{D_1D\psi}}$ }\label{A:3P0}

Here we give an estimate for the ratio of couplings for $D_1DJ/\psi$ and $D_1D\psi^\prime$ by using the $^3P_0$ model~\cite{LeYaouanc:1972ae,*LeYaouanc:1973xz,Barnes:1996ff,Barnes:2005pb,Ackleh:1996yt}. The $^3P_0$ model is a standard phenomenological tool to analyze hadron decays. Thereby a $q \bar q$ pair is created from vacuum with quantum numbers $I^G(J^{PC})=0^+(0^{++})$, hence $^3P_0$ in spin-orbit coupling. The $^3P_0$ model is rather sensitive to variations of the parameters i.e. the elementary pair creation strength and the radii of the hadron wave functions involved. The model can deliver meaningful results for strong hadronic decay rates when evaluated in the center of mass frame and provided that all initial or final state particles are on-shell. Further extensions of the $^3P_0$ model concern for example nucleon-antinucleon annihilation processes~\cite{Dover:1992vj,*Dover:1992jp} and the determination of baryon meson coupling constants ~\cite{Gutsche:1994pk,Barnes:2005pb} where the emitted meson is not necessarily on-shell anymore. Since in the present work we deal with transitions between off-shell mesons, the $D_1\to D\psi$ and $D_1\to D\psi^\prime$ decays are kinematically forbidden, we cannot fix the couplings directly from the model. However,  since the ratio of transition matrix elements has a less pronounced parameter dependence,  we use the $^3P_0$ model to determine the ratio of the couplings $r_2={g_{D_1D\psi^\prime}}/{g_{D_1D\psi}}$. 

In the $^3P_0$ model the transition amplitude for the process $A\to BC$ is given by
\be
T_{A\to BC}=\big<\Psi^B_{n_B,l_B,m_B}(1,3)\Psi^C_{n_C,l_C,m_C}(2,4)\big|\hat{\cal O}_{^3P_0}(3,4)\big|\Psi^A_{n_A,l_A,m_A}(1,2)\big>\,,
%=\int\prod\limits_{i=1,2,1^\prime,..,4^\prime}d^3p_i\psi_{BC}^\dagger\hat {\cal O}_{^3P_0}\psi_A\,.
\en
where the indices $i=1,2,3,4$ refer to the respective quarks. For the initial and final mesons we use, as usual, simple harmonic oscillator wave functions
\bea
\psi_{n,l,m}(\vec p)&=&N_{n,l}L_n^{l+1/2}(R^2p^2)\exp\big(-\frac{R^2}{2}{\vec p}^{\;2}\big)(Rp)^l\big[\big|(s_1s_2)sm_s\big>\otimes{Y}_{lm}(\hat p)\big]_{J,m_{J}}\\\nonumber
&&\times\delta^{(3)}(\vec P-\vec p_1-\vec p_2)\chi^s_m(12)\chi^{f+c}(q_1q_2)\,,
\ena
with the normalization $N_{n,l}=(-i)^{2n+l}\sqrt{\frac{2n!R^3}{\Gamma(n+l+3/2)}}$, the radius $R$ and the relative momentum $\vec p=\frac{m_1\vec p_2-m_2\vec p_1}{m_1+m_2}$. Quark $i$ is characterized by its mass $m_i$ and spin $s_i$. $L_n^l$ and $Y_{lm}$ represent the Legendre and the spherical harmonics, respectively. 
$\chi^s$ and $\chi^{f+c}$ denote the spin and flavor-color wave functions.

The additional quark-antiquark pair required for the decay of a meson into a two-body final state is generated by the $^3P_0$ operator
\bea
\hat {\cal O}_{^3P_0}&=&\lambda{V_{^3P_0}^{\scriptstyle{34}}}^\dagger\underbrace{\delta^{(3)}(\vec p_1-{\vec p_1}^{\,\prime})\delta^{(3)}({\vec p_2-\vec p_2}^\prime)}_{\text{\footnotesize{spectator quarks}}}
\ena
where $\lambda$ is the constant $q\bar q$ production amplitude and 
\bea
{V_{^3P_0}^{\scriptstyle{34}}}^\dagger&=&\sum\limits_\mu(-)^{1+\mu}\big<1\,1\,-\mu\,\mu\,\big|00\big>{\cal Y}_{1\mu}^\ast(\vec p_3-\vec p_4)\delta^{(3)}(\vec p_3-\vec p_4){\sigma_{-\mu}^{(34)}}^\dagger
\ena
with the Pauli matrix $\sigma$ and  ${\cal Y}_{lm}(\vec p)=\big|p\big|Y_{lm}(\hat p)$.

%\subsection{$\textstyle{\mathbf{2(1)^3S_1\to ^1P_1(^3P_1)+^1S_0}\quad}$ ($\textstyle{\psi^{(\prime)}\to D_1D}$)}

In the following we give the final result for the amplitudes characterizing the $\psi^{(\prime)}\to D_1D$ transitions $(2)1^3S_1\to ^1P_1(^3P_1)+^1S_0$, where only the $D$-wave contribution ($L=2$) is relevant
$$
T_{L=2}=\lambda\delta^{(3)}(\vec P_A-\vec P_B-\vec P_C)\sum\limits_{L,m_L}Y_{Lm_L}(\hat P)T_L^{\text{space}}f_L(\mu,m_{l_B},m_L)
$$
with
\beaa
T_{L=2}^{\text{space,n=0}}&=&i\frac{\sqrt R2^3}{\pi^{\frac34}3^2}\sqrt{\frac{1}{5}}\alpha\big(1-\frac23\alpha\big)R^2P^2\exp\big(-\frac{R^2}{3}P^2\alpha^2\big)\,,\\
T_{L=2}^{\text{space,n=1}}&=&-3i\frac{\sqrt R2^{\frac92}}{3^{\frac92}\pi^{\frac34}}\sqrt{\frac{1}{5}}\Big(\frac{11}{4}\alpha-\frac{7}{6}\alpha^2-\frac23\alpha^3R^2P^2+\frac{4}{9}\alpha^4R^2P^2\Big)R^2P^2\exp\big(-\frac{R^2}{3}P^2\alpha^2\big)
\enaa
 and
\eq
f_L(\mu,m_{l_B},m_L)=\big<1\,1\,0\,0\big|2\,0\big>\big<1\,1\,-\mu\,-m_{l_B}\big|L\,m_L\big>(-)^{m_L}\delta_{L,2}\label{eq:l}
\en
(see also in e.g.~\cite{Barnes:1996ff}). Here $L$ is the relative angular momentum between the final mesons and $J_{BC}$ is the total spin of the two final states $B$ and $C$. The different quark masses of the charm mesons are accounted for by the factor $\alpha=\frac{m_c}{m_q+m_c}$.
The latter expression (\ref{eq:l}) represents the spin part of the amplitude for a $^1P_1$ and $^3P_1$ state, respectively
\beaa
T^{\text{spin}}_{^1P_1}&=&\sqrt{\frac23}\frac{\sqrt{10}}{6}\big<J_{BC}\,2\,m_{BC}\,m_L\big|1\,m_A\big>\delta_{L,2}\,,\\
T^{\text{spin}}_{^3P_1}&=&\sqrt{\frac13}\frac{\sqrt{10}}{6}\big<J_{BC}\,2\,m_{BC}\,m_L\big|1\,m_A\big>\delta_{L,2}\,.
\enaa
Finally we consider the mixing $\left|D_1\right>=\sqrt{\frac23}{\left|^1P_1\right>}+\sqrt{\frac13}{\left|^3P_1\right>}$, which leads to 
\eq
T_{\psi\to D_1 D}^{\text{space,n=0(1)+spin}}=\sqrt{\frac32}T^{\text{space,n=0(1)}}_{L=2}\,T^{\text{spin}}_{^1P_1}\,.
\en
Due to this mixing pattern the $D_1$ couples via $D$-wave only.

We find that the ratio $g_{D_1D\psi^\prime}/g_{D_1D\psi}=T^{\text{space,n=1}}_{L=2}/T^{\text{space,n=0}}_{L=2}=2.15$ is independent on the radii $R$ and coupling strength $\lambda$, where we evaluated the amplitudes at threshold i.e. $P$=0 which is in analogy to the determination of the nucleon-meson couplings in~\cite{Downum:2006re}. The quark masses $m_q=0.33$~GeV and $m_c=1.6$~GeV are taken from~\cite{Barnes:2005pb}.
\end{appendix}
\end{document}